\documentclass[twoside]{article}
\usepackage[accepted]{aistats2016}

\ifnum\statePaper=1
    \newcommand{\blind}[1]{#1}
\else
    \newcommand{\blind}[1]{\emph{Anonymized}}
\fi

\usepackage[round]{natbib}
\usepackage{amsmath,amssymb,url,graphicx,microtype,booktabs}
\usepackage{subfig}
\usepackage{bibhacks}
\newcommand{\citeg}[1]{\citep[e.g.][]{#1}}

\newcommand{\te}{\!=\!} 
\newcommand{\tp}{\!+\!} 
\newcommand{\tm}{\!-\!} 
\newcommand{\ttimes}{\!\times\!} 

\newcommand{\g}{\,|\,} 

\newcommand{\bu}{{\bf u}}

\newcommand{\E}{\mathbb{E}}

\newcommand{\N}{\mathcal{N}}

\newcommand{\hf}{\widehat{f}}

\usepackage{multirow}
\usepackage[pdfborder={0 0 0}]{hyperref}

\newcommand{\gvn}{;\,}
\newcommand{\est}{\epsilon}

\newcommand{\state}{\theta}
\newcommand{\statemn}{\state_\mathrm{min}}
\newcommand{\statemx}{\state_\mathrm{max}}
\newcommand{\vct}[1]{\mathbf{#1}}
\newcommand{\mtx}[1]{#1}

\newcommand{\Zconst}{C}
\newcommand{\Zdi}{Z}

\newcommand{\sliceh}{h}
\newcommand{\didata}{y}

\newcommand{\MH}{MH}

\begin{document}

\twocolumn[
\aistatstitle{Pseudo-Marginal Slice Sampling}
\aistatsauthor{Iain Murray \And Matthew M. Graham}
\aistatsaddress{University of Edinburgh \And University of Edinburgh}]

\begin{abstract}
    Markov chain Monte Carlo (MCMC) methods asymptotically sample from
    complex probability distributions.
    The pseudo-marginal MCMC framework only requires an unbiased estimator of the
    unnormalized probability distribution function to construct a Markov chain.
    However, the resulting chains are harder to tune to a target
    distribution than conventional MCMC, and the types of updates available are
    limited. We describe a general way to clamp and update the random numbers
    used in a pseudo-marginal method's unbiased estimator. In this framework we can use slice sampling and
    other adaptive methods.
    We obtain more robust Markov chains, which often mix more quickly.
\end{abstract}

\section{Introduction}

Markov chain Monte Carlo (MCMC) methods asymptotically sample from a
user-specified probability distribution, by simulating a Markov chain with
the required equilibrium distribution.
Most MCMC methods require an ability to evaluate the target distribution up
to a constant. Pseudo-marginal MCMC \citep{andrieu2009} only
requires unbiased estimates of the target distribution.
Pseudo-marginal MCMC has been applied across a range of domains, with
applications in genetic population modelling \citep{beaumont2003},
continuous time stochastic processes \citep{georgoulas2015}, hierarchical
models involving Gaussian processes \citep{filippone2014}, and
`doubly-intractable' distributions \citep{murray2006} such as undirected
graphical models.

As the number and complexity of applications of MCMC increases, methods
that
adapt free parameters for the user are becoming more popular
\citeg{hoffman2014,murray2010,roberts2009}.
Slice sampling \citep{neal2003a} is relatively insensitive to its
free parameters, and can locally adapt to the target density.
Unfortunately,
these methods either can't be applied with
pseudo-marginal MCMC methods, or no longer work as designed.

In this paper we explore a simple means to bring together the advantages of
pseudo-marginal methods and slice sampling. Like recent work on particle
Gibbs methods \citep{chopin2015}, we identify randomness used in an
unbiased estimator as auxiliary variables that could be updated with a
variety of MCMC methods. The framework is simple, and requires only a small
change to existing pseudo-marginal methods that use Metropolis--Hastings
style updates. Where the estimator has high variance, our methods can lead
to a large increase in performance. The framework also allows
Metropolis--Hastings step sizes to be tuned, or
slice sampling to be used, and can make pseudo-marginal chains more robust and less
prone to sticking.
\looseness=-1

\section{Background}

The standard task for Markov chain Monte Carlo (MCMC) is to draw correlated
samples from a target distribution of interest
\begin{equation}
    \pi(\theta) = f(\theta)/\Zconst,
    \label{eqn:target}
\end{equation}
where we need to be able to evaluate $f(\theta)$ pointwise, but not the
normalizer~$\Zconst$.

\subsection{Pseudo-Marginal MCMC}

In this paper we consider cases where $f(\theta)$ cannot easily be evaluated pointwise,
but where an unbiased
estimator $\hf(\theta)$ can be sampled for any given state~$\theta$. This
estimator might come from standard importance sampling estimates, particle
methods \citep{andrieu2010}, or randomized series truncation
\citep{girolami2013, georgoulas2015}.

\begin{figure}%
\vspace*{-0.09in}
\small
\begin{tabular}{l}
\hphantom{\hspace*{0.9\linewidth}}\\
\toprule
\begin{minipage}{0.95\linewidth}
\textbf{Inputs:} current state $\theta$,
previous estimate of its unnormalized target probability $\hf$,
proposal dist.\ $q$,
unbiased estimator s.t.\ $\E_{\est(\hf\gvn\theta)}[\hf] = f(\theta)$ for all $\theta$,

\smallskip

\textbf{Output:} new state-estimate pair $(\theta,\hf)$.
\end{minipage}\\
\midrule
\end{tabular}

\vspace*{-0.2cm}

\begin{enumerate}
\setlength{\itemsep}{1pt}
\setlength{\parskip}{0pt}
\setlength{\parsep}{0pt}
\item Propose new state and estimate its probability:\\[-0.07in]
\begin{equation*}
\begin{split}
         \theta' &\,\sim\, q(\,\cdot\,;\,\theta)\\
    \hf'  &\,\sim\, \est(\,\cdot\,\gvn \theta')
\end{split}
\end{equation*}%
\item Metropolis--Hastings style acceptance rule,\\[-0.07in]
\begin{equation*}
    \mbox{\textbf{with probability}}
    \min\!\bigg(1,\; \frac{\hf'}{\hf}\,\frac{q(\theta;\,\theta')}{q(\theta';\,\theta)}\bigg)\mbox{\textbf{:}}
    \hspace*{5cm}
\end{equation*}
\item[] \qquad Accept: \textbf{return} $(\theta', \hf')$\\[-0.1in]
\item[] \textbf{else:}\\[-0.1in]
\item[] \qquad Reject: \textbf{return} $(\theta, \hf)$
\end{enumerate}

\vspace*{-0.5cm}
\begin{tabular}{l}
\hphantom{\hspace*{0.95\linewidth}}\\
\bottomrule
\end{tabular}

\caption{%
Pseudo-marginal
update as analysed by
\hbox{\citet{andrieu2009}}.
}
\label{alg:pm}
\end{figure}

In the `pseudo-marginal' or `exact-approximate' paradigm \citep{andrieu2009}, we
can simulate the target distribution \eqref{eqn:target} by constructing a Markov
chain on a state pair $(\theta,\hf)$. The state evolves according to the
algorithmic update in Figure~\ref{alg:pm}.
When the estimator is
deterministic and exact --- $\hf(\theta) \te f(\theta)$ for all $\theta$ ---
the update is precisely the Metropolis--Hastings (\MH) algorithm, with
acceptance ratio
\begin{equation}
    a = \frac{\pi(\theta')\, q(\theta\gvn\theta')}{\pi(\theta)\, q(\theta'\gvn\theta)}
      = \frac{f(\theta')\, q(\theta\gvn\theta')}{f(\theta)\, q(\theta'\gvn\theta)}.
    \label{eqn:mh_acceptance_ratio}
\end{equation}
When the
estimator is noisy, the estimate for the current state must be stored and reused
until a move is accepted. That detail is made explicit by incorporating the
estimate $\hf$ into the Markov chain state.
Special cases of the pseudo-marginal algorithm had previously been identified
and explored \citep{kennedy1985,beaumont2003,moller2006}.

Because $\hf$ is part of the Markov chain state, we must keep \emph{both} the
current variables $\theta$ \emph{and} its corresponding estimate $\hf$ if the
\MH\ rule rejects a move. We can't simply replace the noisy function value with
a new draw of the estimator from its distribution $\est(\hf\gvn\theta)$.
When the distribution of the estimator is heavy-tailed, it is hard to accept a change
away from a large $\hf$, and the chain can stop moving for a long time.
Pseudo-marginal algorithms are notorious for this `sticking' behavior.

\subsection{Doubly-Intractable Distributions}
\label{sec:di}

Pseudo-marginal methods can be applied to \emph{doubly-intractable} target
distributions. These arise
when infering the parameters $\theta$ from a data generating process
\begin{equation}
    p(\didata\g\theta) = g(\didata\gvn\theta) \,\big/\, \Zdi(\theta),
    \label{eqn:di_like}
\end{equation}
where $g(\didata\gvn\theta)$ is a function we can evaluate, but $\Zdi(\theta)$ is an
intractable function of the parameters.
Given a prior over parameters $p(\theta)$ and observation~$\didata$,
the target posterior distribution from Bayes' rule is:
\begin{equation}
    \pi(\theta) \,=\, p(\didata\g\theta)\,p(\theta) \big/ p(\didata) \,=\, f(\theta) \big/ \Zconst. \label{eqn:di}
\end{equation}
We can choose to write this target distribution with reference to
a model with fixed parameters $\hat{\theta}$:
\begin{align}
    f(\theta) &= g(y \gvn \theta)\,p(\theta)\,\frac{\Zdi(\hat{\theta})}{\Zdi(\theta)}, \text{~with unbiased estimator} \notag\\
    \hf(\theta) &= g(y \gvn \theta)\,p(\theta)\,
    \frac{g(x;\hat{\theta})}{g(x;\theta)}, \quad x\sim p(\didata\g\theta).
    \label{eqn:estdi}
\end{align}
This importance sampling estimator requires a sample from each model
considered. Formally, exact/perfect sampling
methods such as `coupling from the past' \citep{propp1996} are required.
Using estimator \eqref{eqn:estdi} in the pseudo-marginal framework
corresponds to the MCMC method proposed by \cite{moller2006}.

\subsection{Slice Sampling}

Slice sampling \citep{neal2003a} is a family of algorithms
with update mechanisms that can locally adapt to the target density.
The Markov chain explores the uniform distribution underneath the $f(\theta)$
surface, so that the probability of being above setting $\theta$ is proportional
to $\pi(\theta)$ as desired. A state/height pair $(\theta,\sliceh)$ are
updated alternately according to Figure~\ref{alg:slice}. \citet{neal2003a}
proved that this procedure (with some conditions) will sample from the target
distribution.

The algorithm in Figure~\ref{alg:slice} has a step-size parameter $w$, but can
adapt to bad settings.
If the step-size is too small, step~\ref{alg_item:lin_step_out} of the algorithm
can expand the interval explored. If the interval is too large, an adaptive
rejection procedure in steps
\ref{alg_item:lin_slice_loop_point}--\ref{alg_item:lin_slice_loop_back} will
shrink the interval exponentially quickly towards the current state until an
acceptable update is found. Given a continuous function $f(\theta)$, the update
always moves the current state.
\looseness=-1

\begin{figure}
\vspace*{-0.09in}
\small
\begin{tabular}{l}
\hphantom{\hspace*{0.9\linewidth}}\\
\toprule
\begin{minipage}{0.95\linewidth}
\textbf{Input:} current state $\state$,
unnormalized target distribution $f$,
initial search width $w$,
whether to do optional part of update \texttt{step\_out}.

\smallskip

\textbf{Output:} a new state $\state'$. When $\state$ is drawn
from $\pi(\state) \!\propto\! f(\state)$, the marginal
distribution of $\state'$ is also~$\pi$.
\end{minipage}\\
\midrule
\end{tabular}

\vspace*{-0.2cm}

\begin{enumerate}
\setlength{\itemsep}{1pt}
\setlength{\parskip}{0pt}
\setlength{\parsep}{0pt}
\item Random height under curve:\\[-0.07in]
\begin{equation*}
\begin{split}
         u_1  &\kern2pt\sim   \mathrm{Uniform}[0, 1]\\
         \sliceh  &\leftarrow  u_1 f(\state)\\
\end{split}
\end{equation*}\\[-0.23in]
\label{alg:lin_slice_height}
\item Randomly place interval around the current state:\\[-0.07in]
\begin{equation*}
\begin{split}
    u_2 &\kern2pt\sim \mathrm{Uniform}[0,w]\\
    [\statemn,\,\statemx] &\leftarrow [\state \tm u_2,\; \state \tm u_2 \tp w]\\
\end{split}
\end{equation*}\\[-0.23in]
\label{alg_item:lin_rand_place_intvl}
\item \textbf{if} \texttt{step\_out} expand interval (linear step version): \label{alg_item:lin_step_out}\\[-0.1in]
\item[] \qquad\textbf{while} $f(\statemn)\kern0.5pt > \sliceh$ \textbf{:}~ $\statemn \kern1pt\leftarrow\kern0.75pt \statemn - w$
\item[] \qquad\textbf{while} $f(\statemx)\kern-1pt > \sliceh$ \textbf{:}~ $\statemx \leftarrow \statemx + w$\\[-0.07in]
\item Sample proposal on interval: \label{alg_item:lin_slice_loop_point}\\[-0.07in]
\begin{equation*}
    \state' \sim \mathrm{Uniform}[\statemn,\,\statemx]
\end{equation*}\\[-0.33in]
\label{alg_item:lin_sample_prop}
\item \textbf{if} $f(\state') \leq \sliceh$ \textbf{then:}\\[-0.1in]
\item[] \qquad Shrink the bracket and try a new point:
\item \qquad\textbf{if} $\state' < \state$ \textbf{then:}
    $\statemn \leftarrow \state'$ \textbf{else:} $\statemx \leftarrow \state'$
    \label{alg_item:lin_slice_shrink}
\item \qquad\textbf{GoTo} \ref{alg_item:lin_slice_loop_point}.
\label{alg_item:lin_slice_loop_back}
\item \textbf{else:} \label{alg_item:lin_branch_in_slice}
\item \qquad Accept: \textbf{return} $\state'$
\label{alg_item:lin_accept_proposal}
\end{enumerate}

\vspace*{-0.5cm}
\begin{tabular}{l}
\hphantom{\hspace*{0.95\linewidth}}\\
\bottomrule
\end{tabular}

\caption{%
Single-variable slice sampling with linear stepping out, adapted from \citet{neal2003a}.
}
\label{alg:slice}
\end{figure}

\section{Noisy Slice Sampling}

As slice sampling algorithms like Figure~\ref{alg:slice} always move the
variables being updated, it is tempting to apply them in the pseudo-marginal
setting, where chains often stick.
It turns out that slice sampling is still valid if replacing the function
$f(\theta)$, with a noisy but unbiased estimate $\hf(\theta)$.
A special case
is discussed by \citet[\S5.8.2]{murray2007}. Intuitively, exploring uniformly
underneath the noisy surface means spending time in a region above $\theta$
proportional to the average value of $\hf(\theta)$, and thus sampling from the
target distribution~$\pi(\theta)$.

Although slice sampling with noisy values can be valid, the local adaptation
of the proposal interval is not designed for this use case. If $f(\theta)$ is
occasionally over-estimated by a large amount, even nearby and equally probable states will be deemed
unacceptable.
In practice the proposal interval can easily
collapse to the current state to numerical machine precision. Although the
variables of interest $\state$ are no longer changing, an implementation must be
careful to either keep proposing $(\state,\hf)$ states until an acceptable $\hf$
has been found, or to return the original $(\state,\hf)$ pair from before the
update began. Either way,
the Markov chain
will effectively contain rejections.

\section{Clamping randomness}

We aim to ease exploration of
variables $\theta$ while exploiting a noisy estimator~$\hf$, by
removing the noise from the update.
We will assume as
little about the estimators as possible, to keep the appealing `black-box' spirit of the
original pseudo-marginal framework.

We assume that the estimation procedure uses
random numbers from a
convenient distribution, $q(\bu)$, such as a uniform or Gaussian distribution.
This assumption is not strong:
the choices
$\bu$ could be the results of
all of the calls to a random number generator (often called \texttt{rand()})
within the computer code for the estimator. The estimate $\hf(\theta\gvn\bu)$ is then
deterministically computed for the current variables given these choices.

Instead of sampling $\bu$ from a random number generator, we will
evolve them as part of a Markov chain on a joint auxiliary target
distribution:
\begin{equation}
    \pi(\theta,\bu)
                    = \hf(\theta\gvn\bu)\,q(\bu) \,/\, \Zconst.
    \label{eqn:jointauxtarget}
\end{equation}
If the estimator
is unbiased, we know its average under its random choices~$\bu$:
\begin{equation}
    \E_{q(\bu)}\kern-1pt\big[\hf\kern2pt\big]
    = \int \hf(\theta\gvn\bu)\,q(\bu) \;\mathrm{d}\bu
    = f(\theta).
\end{equation}
Then the marginal distribution over the variables of interest,
$\pi(\theta)\te \int \pi(\theta,\bu)\,\mathrm{d}\bu\te f(\theta)/\Zconst$, is the
user-specified target marginal distribution~\eqref{eqn:target}.

Pseudo-marginal MCMC can be seen as a standard Metropolis--Hastings update on the auxiliary
target distribution~\eqref{eqn:jointauxtarget}.
After proposing new variables with probability
$q(\theta'\gvn\theta)$ we also propose new random choices with probability
$q(\bu')$
giving joint proposal probability
\begin{eqnarray}
    q(\theta',\bu'\gvn\theta,\bu) = q(\theta'\gvn\theta)\,q(\bu').
    \label{eqn:jointauxprop}
\end{eqnarray}
Substituting \eqref{eqn:jointauxtarget} and \eqref{eqn:jointauxprop} into the
\MH\ acceptance ratio \eqref{eqn:mh_acceptance_ratio} gives
\begin{equation}
    a \,=\, \frac{\pi(\theta',\bu')\,q(\theta,\bu\gvn\theta',\bu')}%
             {\pi(\theta,\bu)\,q(\theta',\bu'\gvn\theta,\bu)}
             \,=\, \frac{\hf(\theta'\gvn\bu')\,q(\theta\gvn\theta')}{\hf(\theta\gvn\bu)\,q(\theta'\gvn\theta)},
    \label{eqn:auxaccept}
\end{equation}
as used in the acceptance rule in Figure~\ref{alg:pm}.

\subsection{Alternative transition operators}

Splitting out the random choices $\bu$ allows us to apply Markov
chain operators that were not available before.
Given a joint distribution $\pi(\bu,\theta)$, we can alternately update
$\bu$ and
$\theta$ using any standard MCMC
updates for the conditional target distributions $\pi(\bu\g \theta)$
and $\pi(\theta\g\bu)$. This is our simple \emph{Auxiliary Pseudo-Marginal} (APM)
framework, Figure~\ref{alg:apm}.

\begin{figure}
\vspace*{-0.09in}
\small
\begin{tabular}{l}
\hphantom{\hspace*{0.9\linewidth}}\\
\toprule
\begin{minipage}{0.95\linewidth}
\textbf{Inputs:} current state: parameters $\theta$, randomness $\bu$;
unbiased estimator s.t.\ $\E_{q(\bu')}[\hf(\theta';\bu')] = f(\theta')$ for all $\theta'$,

\smallskip

\textbf{Output:} new state $(\theta,\bu)$.
\end{minipage}\\
\midrule
\end{tabular}

\vspace*{-0.2cm}

\begin{enumerate}
\setlength{\itemsep}{1pt}
\setlength{\parskip}{0pt}
\setlength{\parsep}{0pt}
\item Update $\bu$ leaving invariant its target conditional:\\[-0.07in]
\begin{equation*}
         \pi(\bu\g\theta) \,\propto\, \hf(\theta; \bu)\,q(\bu)
\end{equation*}%
\item Update $\theta$ leaving invariant its target conditional:\\[-0.07in]
\begin{equation*}
        \pi(\theta\g \bu) \,\propto\, \hf(\theta; \bu)\hphantom{\,q(\bu)}
\end{equation*}%
\end{enumerate}

\vspace*{-0.5cm}
\begin{tabular}{l}
\hphantom{\hspace*{0.95\linewidth}}\\
\bottomrule
\end{tabular}

\caption{%
Framework for Auxiliary Pseudo-Marginal (APM) methods. Here `leaving invariant' means
that if a variable was drawn from the specified probability distribution before
an update, it will retain the same distribution after the update. This
property is satisfied by all standard MCMC update rules.}
\label{alg:apm}
\end{figure}

We could implement the first step in Figure~\ref{alg:apm} by a Metropolis Independence (MI) proposal. We
propose new random choices with probability $q(\bu')$, and accept or reject them
with the standard Metropolis ratio, which here simplifies to
$\hf(\theta;\bu')/\hf(\theta;\bu)$. This step is a standard pseudo-marginal update,
but where the proposal for variables $\theta$ has zero-width.

The difference when updating the variables $\theta$ in the second step of
Figure~\ref{alg:apm} is that
the random choices $\bu$ are fixed. This step could be implemented by simply
resetting the seed of a pseudo-random number generator to the
seed used to produce the estimate that was last accepted in
step~1. Given clamped random numbers, we have the usual case of a conditional
distribution proportional to a deterministic function. Slice sampling
will move $\theta$ if $\hf(\theta;\bu)$ is continuous almost everywhere.

Although we can now ensure that the variables of interest $\theta$ will almost
always move, we might still suffer from sticking of the random choices~$\bu$.
Rather than making global, Metropolis Independence (MI) proposals from $q(\bu)$,
we could instead attempt to perturb~$\bu$ using other MCMC update rules.

\textbf{Naming scheme:~}
In the rest of the paper we will consider Auxiliary Pseudo-Marginal (APM)
methods with a
variety of MCMC algorithms providing the updates for the two steps in
Figure~\ref{alg:apm}. We will use the shorthand MI for Metropolis Independence
updates, MH for perturbative Metropolis--Hastings updates, and SS for slice
sampling.
We will specify the update of the random choices $\bu$ first,
followed by the update for~$\theta$.
For example, an
APM
method which uses a Metropolis
Independence update for $\bu$ given $\theta$, and a slice sampling update for
$\theta$ given $\bu$, will be denoted APM MI+SS\@.
\looseness=-1

\subsection{Slice Sampling auxiliary $\bu$ variables}
\label{sec:ssu}

As discussed above, we recommend trying perturbative proposals of the random
choices $\bu$. In this paper we will only consider methods with no tweak
parameters, as we want our algorithms to be easy to use.

We will use \emph{Elliptical Slice Sampling}
\citep{murray2010} if $q(\bu)$ is Gaussian. This algorithm has no free parameters, and only requires
samples from $q(\bu)$ and the ability to evaluate $\hf(\theta;\bu)$. The
algorithm initially attempts to make large, nearly-independent moves and should
work well if the original Metropolis Independence updates do. Like
conventional slice sampling, the elliptical version can also back off to
smaller moves exponentially quickly.

In many implementations of estimators, the initial random choices $\bu$ will be
uniformly distributed.
Here we use a variant of linear slice sampling (like Figure~\ref{alg:slice}) but
applied along a random direction. We sample a direction vector $\nu$ (with
a random length) by sampling elements independently from a standard normal,
$\nu_i \sim \N(0,1)$. The slice sampler explores an interval aligned with this
direction, and reflects off the unit hypercube boundaries.
Our implementation updates a variable $z$, which
defines the new setting of the randomness:
\begin{equation}
    u'_i = \mathrm{Reflect}(u_i + z\nu_i),
    \label{eqn:newu}
\end{equation}
where we define $m = \mathrm{mod}(x, 2)$ and then
\begin{equation}
    \mathrm{Reflect}(x) = \begin{cases} m & m<1\\ 2-m &m\ge1\end{cases}.
\end{equation}
Each update starts with $z\te0$. We fix the step width to $w\te1$ and omit
the stepping out (Figure~\ref{alg:slice},
step~\ref{alg_item:lin_step_out}). The slice sampling routine needs to evaluate
the target density proportional to $\hf(\theta;\bu'(z))$. If the estimator
function $\hf$ chooses the number of random numbers it uses on the fly, these
can be lazily generated from \eqref{eqn:newu}. Each $\nu_i$ is an independent
Gaussian variate. Each $u_i$ is either a cached random number that was
previously used, or it didn't affect the previous estimator outcome, and can
be retrospectively sampled from a uniform distribution.
\looseness=-1

Because reflections have unit Jacobian, the original proofs for slice sampling
follow through. Reflections in slice sampling have previously been considered by
\citet{downs2000} and \citet{neal2003a}.

\begin{figure*}
    \vspace*{-0.09in}
    \hspace*{-0.1cm}
    \subfloat[][PM-MH~~~~]{\includegraphics[scale=0.55]{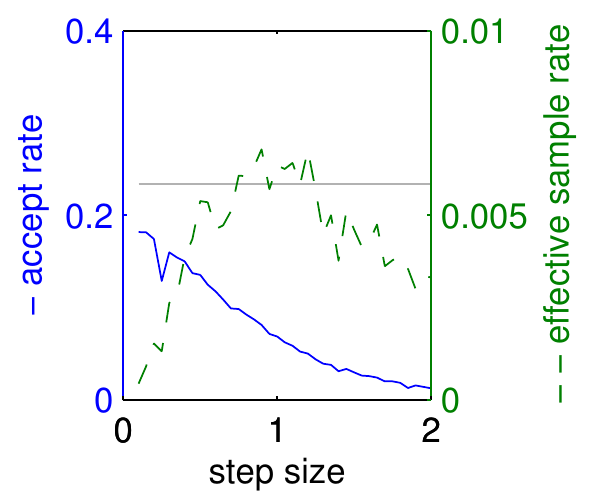}~\label{fig:pmar}}
    \hspace*\fill
    \subfloat[][APM MI+MH~~]{\includegraphics[scale=0.55]{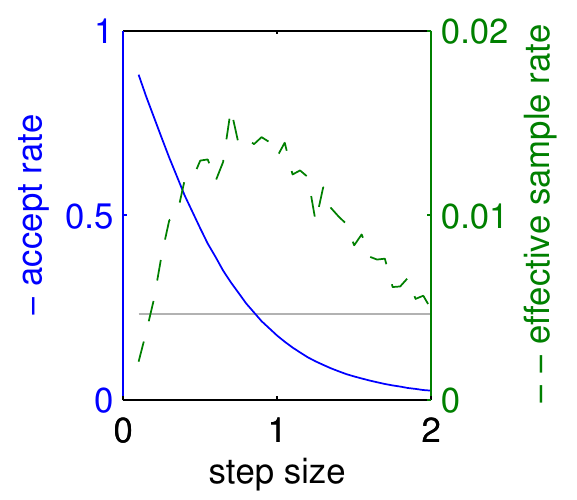}\label{fig:pmcar}}
    \hspace*\fill
    \subfloat[][$p(\theta_1)$ estimates]{\includegraphics[scale=0.55]{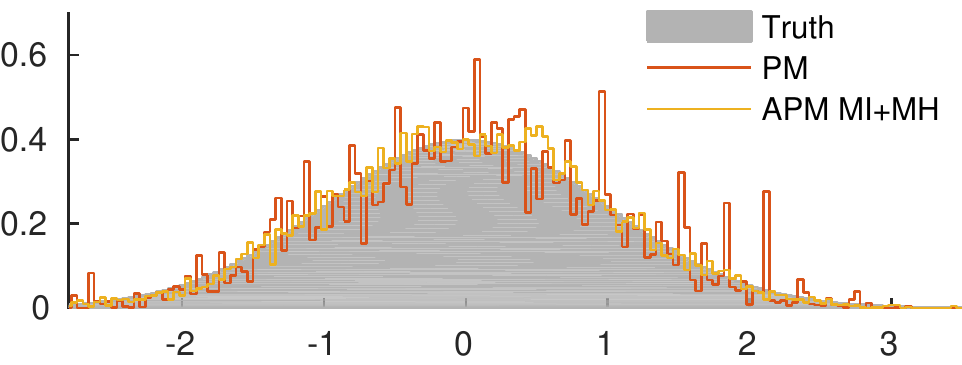}\label{fig:toyhist}}
    \hspace*\fill
    \subfloat[][Cost-scaled autocorrelation]{\includegraphics[scale=0.55]{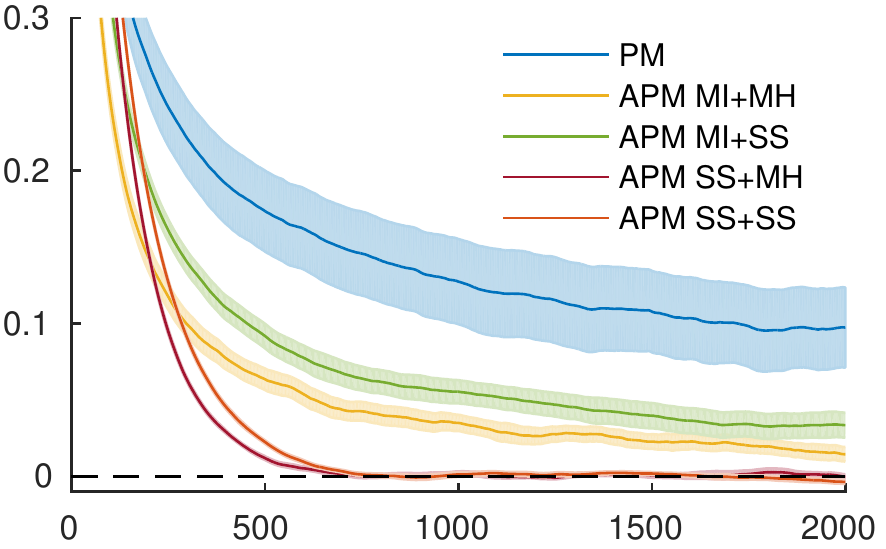}~\label{fig:toycorr}}
    \caption{Sampling from a 5-dimensional Gaussian distribution, as if it were
    doubly-intractable. Acceptance rate and effective sample size per iteration for
    \protect\subref{fig:pmar}~pseudo-marginal \MH\ updates (PM-MH), and
    \protect\subref{fig:pmcar}~updates split into two steps,
    where randomness $\bu$ in the estimator is `clamped' while updating variables
    $\theta$ (APM MI+MH). The gray horizontal line, an acceptance rate of
    $0.234$, indicated an \MH\ step-size of 0.85 for further runs.
    \protect\subref{fig:toyhist}~Estimates of a marginal distribution from
    100,000 PM-MH updates, and 50,000 APM MI+MH update pairs.
    \protect\subref{fig:toycorr}~Autocorrelation plotted against
    steps taken multiplied by the relative cost per step compared to standard PM
    updates.
    }
\end{figure*}

\section{Demonstrations}

As a first illustration, we choose a 5-dimensional Gaussian
target distribution, $\pi(\theta) = \N(\theta\gvn 0, I)$. We can imagine
this distribution results from a doubly-intractable setup
(Section~\ref{sec:di}) with:
\begin{equation}
    g(x;\theta) = \N(x;\theta, I), \quad y=0, \quad p(\theta) = 1~\text{(improper)},  \notag
\end{equation}
We pretend the normalizer $\Zdi(\theta)=\sqrt{2\pi}$ is unknown, and use the
estimator \eqref{eqn:estdi} with reference parameter $\hat{\theta}\te0$. The
estimator's Gaussian variates $x$, are created from standard normal draws in the
usual way:
\begin{equation}
    \bu \sim \N(0,I), \quad x(\bu\gvn\theta) = \bu + \theta,
    \label{eqn:toyu}
\end{equation}
so our methods use $q(\bu) = \N(\bu\gvn0,I)$.

Updates to the target variables $\theta$ were proposed from a spherical Gaussian
with step size $\sigma$, $q(\theta'\gvn\theta) = \N(0,\sigma^2I)$, or by
one-dimensional slice sampling along a random direction (chosen by finding the
direction of a random draw from $\N(0,I)$) with step-size $w\te4$ and no
stepping out.

\textbf{Clearer step-size selection:}
The optimal step-size (or proposal standard deviation) for
standard Metropolis proposals in high-dimensions (but not when updating one
variable at a time) is the one that gives an acceptance rate of $0.234$ under
quite general conditions \citep{roberts1997}. The usual pseudo-marginal
algorithm, does not meet these conditions however. In fact,
an acceptance rate of $0.234$ isn't achievable
in our Gaussian test case, even for small step sizes
(Figure~\ref{fig:pmar}).
Thus step sizes for pseudo-marginal \MH\ can't be tuned with the same heuristics as standard \MH\@.

Splitting each Markov chain update into two steps, as in Figure~\ref{alg:apm},
means we can perform conventional \MH\ proposals on the variables~$\theta$. We
use independent proposals for randomness $\bu$, by simply re-running our
estimator code as usual without moving the variables~$\theta$. We then make
Gaussian proposals for variables $\theta$ as before, while leaving $\bu$ fixed.
Figure~\ref{fig:pmcar} shows the acceptance rate for just the updates of
variables $\theta$, as a function of their step-size. As expected, the optimal
acceptance rate now appears to be close to the theoretical value of $0.234$.

\textbf{Efficiency:~}
It appears that the effective sample size (ESS) per update \citep{plummer2006} is slightly more than
double what it was before. That implies a similar efficiency to before, given that each APM
MI+MH update contains two evaluations of the estimator. However, ESS estimates are
misleading on poorly mixing chains. Given the same amount of computation, APM
MI+MH gives a smoother estimate of a marginal distribution, while PM still contains
artifacts due to long-lasting sticking of its chain (Figure~\ref{fig:toyhist}).

\textbf{Slice sampling:~} Naively running standard slice sampling using the
noisy function $\hf(\theta)$ worked very poorly, with each update using $>\!100$
function evaluations on average. This idea was abandoned.
Splitting up the $\bu$ and $\theta$ updates allowed slice
sampling the $\theta$ variables to work, although slightly less well than MH
after taking computation cost into account (Figure~\ref{fig:toycorr}). Elliptical slice sampling updates of
the $\bu$ variables gave large improvements in the autocorrelation of the chain,
even when adjusting for the extra computation over the standard independent pseudo-marginal
updates.

\subsection{Ising model parameter posterior}
Following previous studies \citep{moller2006,murray2006},
our second illustration is an Ising model distribution with
$\didata_i\!\in\! \{\pm 1\}$ on a graph with nodes $i$ and edges $E$:
\begin{equation}
    p(\didata\g\theta) = \frac{1}{\Zdi(\theta)} \exp\Big( \sum_{i\neq j \in E} \theta_J \didata_i \didata_j + \sum_i \theta_h \didata_i \Big) .
    \label{eqn:ising}
\end{equation}
Our experiments used a $10\!\times\! 30$ toroidal square Ising lattice.
The data $\didata$ were generated from an exact sample with $\theta_J=0.3$ and
$\theta_h=0$.
We used uniform priors over $|\theta_h|<1$ and $0<\theta_J<0.4$, and an
\MH\ step-size of $\sigma\te0.04$.

We used the {\em Summary States} algorithm \citep{childs2001}
to draw exact samples for the estimator in equation~\eqref{eqn:estdi}.
The reference distribution $\hat\theta$ was set to the true parameters.
Here we identify $\bu$ as the infinite sequence of uniform random numbers used
by a Gibbs sampler started infinitely long ago, which produces an
exact sample from \eqref{eqn:ising}. The summary
states algorithm lazily requests a finite subset of these numbers as required.
When perturbing $\bu$, these were provided on demand by reflective
slice sampling (Section~\ref{sec:ssu}).

We also considered `MAVM' \citep{murray2006}, which replaces the simple
importance sampling in \eqref{eqn:estdi} with an annealed importance sampling
(AIS) estimate \citep{neal2001}. Here the random choices $\bu$ consists of a
fixed length of uniform draws required by the annealing steps of the algorithm,
followed by a variable-length sequence required for an initial exact sample as
before.

Updates to the parameters $\theta_h$ and $\theta_J$ were applied sequentially
using one-dimensional Gaussian proposals or slice sampling updates. Slice
sampling used $w\te0.1$ (not carefully tuned) and no stepping out.
\begin{figure}[t!]
    \vspace*{-0.09in}
    \hspace*{-0.1cm}
    \hfill\subfloat[][PM MH]{\includegraphics[width=0.4\linewidth]{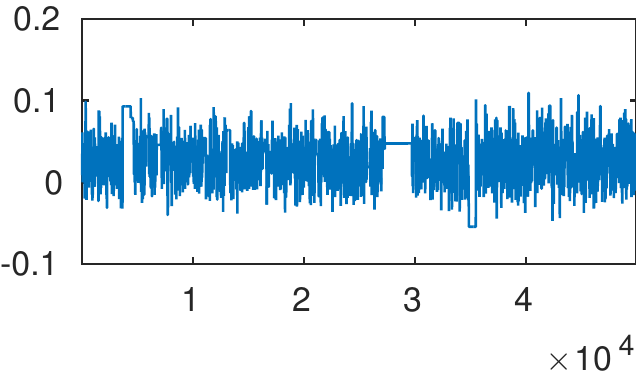}~\label{fig:trace_pm}}%
    \hfill\subfloat[][APM MI+MH]{\includegraphics[width=0.4\linewidth]{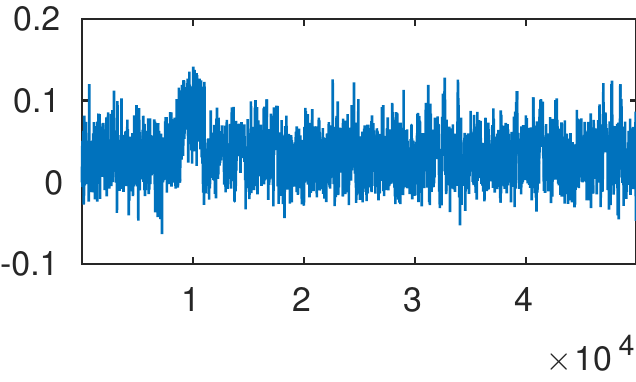}~\label{fig:trace_pm_i_mh}}\hspace*\fill\\[-0.1in]
    \hspace*\fill~~\subfloat[][APM MI+SS]{\includegraphics[width=0.4\linewidth]{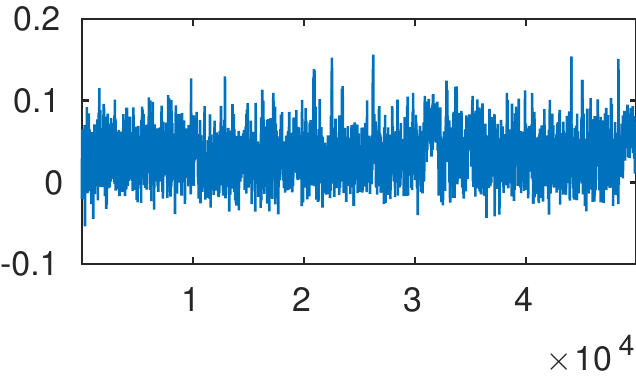}~\label{fig:trace_pm_i_ss}}\hfill%
    \hfill\subfloat[][APM SS+SS]{\includegraphics[width=0.4\linewidth]{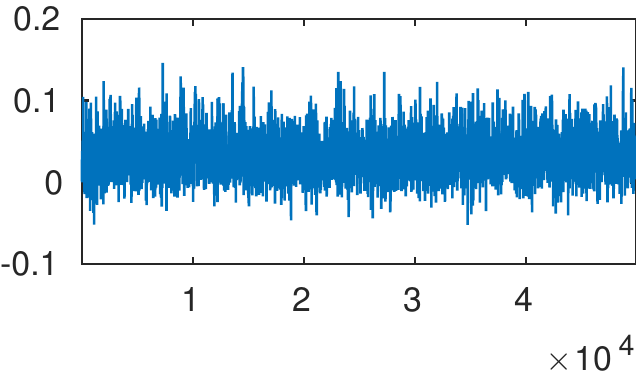}~\label{fig:trace_pm_ss_ss}}\hspace*\fill~~\\[-0.1in]
    \hspace*\fill\subfloat[][Cost-scaled autocorrelations]{\includegraphics[width=0.9\linewidth]{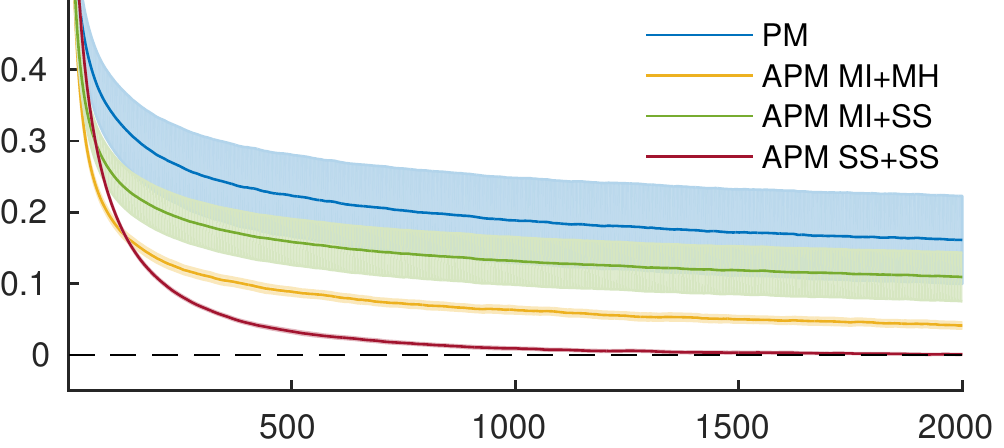}~\label{fig:isingcorr}}\hspace*\fill\\[-0.1in]
    \hspace*\fill\subfloat[][Cost-scaled autocorrelations, with $K\te35$ AIS steps]{\includegraphics[width=0.9\linewidth]{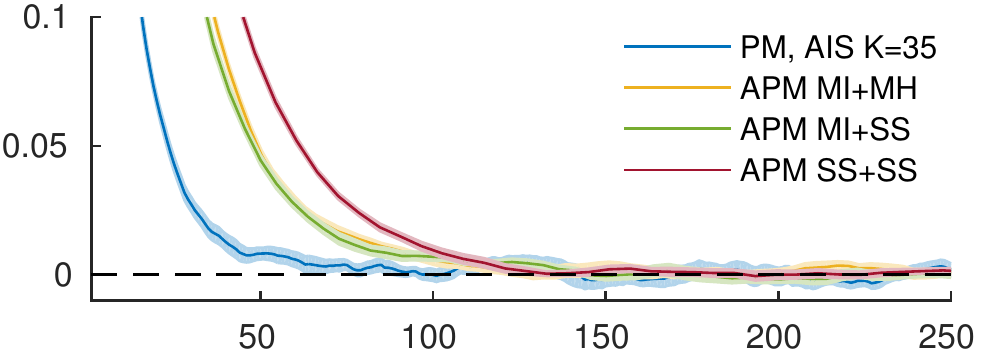}~\label{fig:isingcorrK}}\hspace*\fill
    \caption{Markov chains on the Ising model posterior.
    \protect\subref{fig:trace_pm}~Trace plot of $\theta_h$ against update number
    for the standard pseudo-marginal algorithm.
    \protect\subref{fig:trace_pm_i_mh}--\protect\subref{fig:trace_pm_ss_ss}~Traces
    from versions with separate Independent proposals (I), Gaussian Metropolis
    proposals (MH) or slice sampling updates (SS) for $\bu$+$\theta$.
    \protect\subref{fig:isingcorr}~Empirical cost-scaled auto-correlations based on 10 runs
    of 2~million iterations each. The APM curves are stretched horizontally
    relative to the extra computation performed per iteration compared to the
    original algorithm.
    \protect\subref{fig:isingcorrK}~Empirical auto-correlations when
    using an annealed importance sampling based estimator with $K\te35$ steps. The $x$-axis
    is scaled relative to the costs of one standard update without annealing.
    }
\end{figure}

\textbf{Efficiency:~} The Ising model example showed similar behavior to the toy Gaussian example. A trace plot of the first
$5\ttimes10^4$ updates
(Figure~\ref{fig:trace_pm}) shows that the pseudo-marginal algorithm stuck on
one occasion for $>2000$ iterations. When $\bu$ is updated with separate
independent proposals, the parameters can move at each step. Under these updates
(MI+MH and MI+SS) the auxiliary variables $\bu$ still stick, with a
noticeable effect on $\theta$
(Figures~\ref{fig:trace_pm_i_mh},~\ref{fig:trace_pm_i_ss}). In contrast, slice
sampling $\bu$ forces it to move. The trace plot
(Figure~\ref{fig:trace_pm_ss_ss}) and autocorrelations
(Figure~\ref{fig:isingcorr}, adjusted for computational cost) appear greatly
improved. The consistent behaviour of the SS+SS chain means that the empirical
error bands on its autocorrelation are too small to see.

The empirical autocorrelations (Figure~\ref{fig:isingcorr}) illustrate a danger
of MCMC methods. It appears that the MI+MH updates have a
tightly-determined autocorrelation compared to the more expensive MI+SS updates, from the same number of
iterations. However, the poorer performance and large error band
for MI+SS results from one extended transient, a change of
behavior for $>500,000$ iterations at the end of one run.
As APM MI+MH shares the same independent $\bu$ updates that cause the
sticking, its performance is almost certainly over-stated, and longer runs would
reveal that its autocorrelation is worse
than reported.

The Markov chains can be improved by improving the variance of the
estimator. \citet{murray2006} reported only a modest improvement in effective
sample size of $\sim50\%$ by replacing importance sampling with annealed
importance sampling (AIS)\@. However, given the poor convergence we have
observed without annealing, effective sample size estimates are not reliable.
Figure~\ref{fig:isingcorrK} indicates that the autocorrelations with AIS are in
fact dramatically better (note the different $x$-axis ranges). Applying our
framework to this improved Markov chain gives Markov chains with still faster
convergence per iteration. However, when scaling the autocorrelations by compute
cost, as in Figure~\ref{fig:isingcorrK}, the original MAVM algorithm appears to
be better than our proposed variants.
\looseness=-1

Most MCMC methods don't give guarantees of when estimates will reliably
converge. Re-running APM SS+SS without annealing for 100 million further
updates still didn't reveal any problems, which shows it is less prone to
sticking than the base algorithm, but guarantees little. However, the marginal
distributions show very close agreement to those found with annealing
methods.

\subsection{Gaussian process parameter inference}

\begin{table*}[t!]
\renewcommand{\arraystretch}{1.3}
\small
\begin{tabular}{c|c|cc|ccc|ccc}
 & & & & \multicolumn{3}{c|}{Variance $\sigma$}
 & \multicolumn{3}{c}{Length scale $\tau$} \\
 & Method & $N_{\text{c.op.}} /\,10^3$ & Acc. rate
 & ESS &
   $\frac{\text{ESS}}{N_\text{c.op.}} /\, 10^{-3}$ & $\hat{R}$
 & ESS &
   $\frac{\text{ESS}}{N_\text{c.op.}} /\, 10^{-3}$ & $\hat{R}$ \\
 \hline
 \parbox[t]{2mm}{\multirow{3}{*}{\rotatebox[origin=c]{90}{Pima}}}
 & \sc pm mh & 74.0 (0.026) & 0.201 (0.0065)
 & 306 (7.8) & 4.14 (0.11) & 1.00
 & 441 (8.6) & 5.96 (0.12) & 1.00 \\
 & \sc apm mi+mh & 74.1 (0.029) & 0.219 (0.0034)
 & 357 (8.8) & 4.81 (0.12) & 1.00
 & 512 (19) & 6.92 (0.26) & 1.00 \\
 & \sc apm ss+mh & 74.1 (0.028) & 0.204 (0.0046)
 & 370 (7.1) & 4.99 (0.097) & 1.00
 & 526 (26) & 7.11 (0.35) & 1.00 \\
 \hline
 \parbox[t]{2mm}{\multirow{3}{*}{\rotatebox[origin=c]{90}{Breast}}}
 & \sc pm mh & 97.8 (0.14) & 0.180 (0.013)
 & 185 (19) & 1.90 (0.20) & 1.01
 & 277 (28) & 2.83 (0.28) & 1.00 \\
 & \sc apm mi+mh & 98.3 (0.060) & 0.208 (0.0046)
 & 533 (5.8) & 5.43 (0.059) & 1.00
 & 559 (13) & 5.69 (0.13) & 1.00 \\
 & \sc apm ss+mh & 98.4 (0.054) & 0.206 (0.0044)
 & 519 (9.8) & 5.27 (0.099) & 1.00
 & 631 (13) & 6.41 (0.13) & 1.00 \\
 \hline
\end{tabular}
\caption{Convergence and efficiency results for Gaussian process parameter
inference on two UCI classification data sets. All figures are means (standard
error estimates given in parenthesis) across 10 independent chains initialised
from the prior except for the Gelman--Rubin $\hat{R}$ statistic which is
calculated across all chains. Effective sample sizes ESS are shown
for both variance hyperparameter $\sigma$ and length scale $\tau$ as well as
value normalised by the total number of $\mathcal{O}(n^3)$ matrix operations
computed during each run, $N_\text{c.op.}$ for a measure of overall
computational efficiency.}
\label{tab:gp_param_inf_stats}
\end{table*}

Our third illustration is a
hierarchical Gaussian process classification model, following \citet{filippone2014}. The target distribution is
the posterior
$p(\theta\g\vct{y}, \mtx{X}) \propto p(\vct{y}\g\theta,\mtx{X})\,p(\theta)$
on the model parameters $\theta$ given a set of $n$ observed input features
$\mtx{X} = \left\lbrace \vct{x}_1 ~\dots~\vct{x}_n\right\rbrace$ of dimension
$d$ and a corresponding vector $\vct{y}$ of $n$ binary targets
$y_i \in \lbrace -1,+1 \rbrace$.

The latent function values $\vct{f}$ have a
probit likelihood
$p(\vct{y}\g\vct{f}) = \prod_{i=1}^n \Phi(y_i f_i)$, and
a zero-mean Gaussian process prior
$p(\vct{f}\g\theta, \mtx{X}) = \mathcal{GP}(\vct{f}\g\vct{0},\mtx{K})$ with an
isotropic squared exponential covariance function.
The parameters $\theta = (\sigma,\tau)$ are the variance $\sigma$ and length-scale
$\tau$ of the covariance.
Both $\sigma$ and $\tau$ have Gamma priors.

The marginal likelihood term $p(\vct{y}\g\theta,\mtx{X})$ in the target
posterior cannot be evaluated exactly as the integral to marginalise out the
latent function values
does not have a closed form solution.
\citet{filippone2014} propose using pseudo-marginal MH to sample
from the target posterior. An importance sampling estimate
\begin{equation}
    \begin{split}
    \hat{p}(\vct{y}\g\theta,\mtx{X}) =
    \frac{1}{N_\text{imp}}
    \sum_{i=0}^{N_\text{imp}} \left[
        \frac{p(\vct{y}\g\vct{f}_i)\,p(\vct{f}_i\g\theta,\mtx{X})}
             {q(\vct{f}_i\g\vct{y},\theta,\mtx{X})}
    \right],
    \\
    \vct{f}_i \sim q(\cdot\g\vct{y},\theta,\mtx{X}),
    \end{split}
    \label{eqn:gp_marg_lik_estimator}
\end{equation}
where $q(\vct{f}\g\vct{y},\theta,\mtx{X})$ is a Gaussian approximation to the
posterior on the latent function values, is used as an unbiased estimator of
$p(\vct{y}\g\theta,\mtx{X})$.

We compared the performance of the pseudo-marginal MH
algorithm to two auxiliary pseudo-marginal variants: MI+MH and SS+MH\@. Here
$\bu$ was the fixed-sized vector of standard normal random draws that is
transformed to
generate samples from the Gaussian approximate posterior
$q(\vct{f}\g\vct{y},\theta,\mtx{X})$. We used elliptical slice sampling to update $\bu$ in APM SS+MH\@.

\begin{figure}[t!]
\vspace*{-0.09in}
\hspace*\fill\subfloat[][$\tau$ given Pima data]{\includegraphics[height=4.0cm]{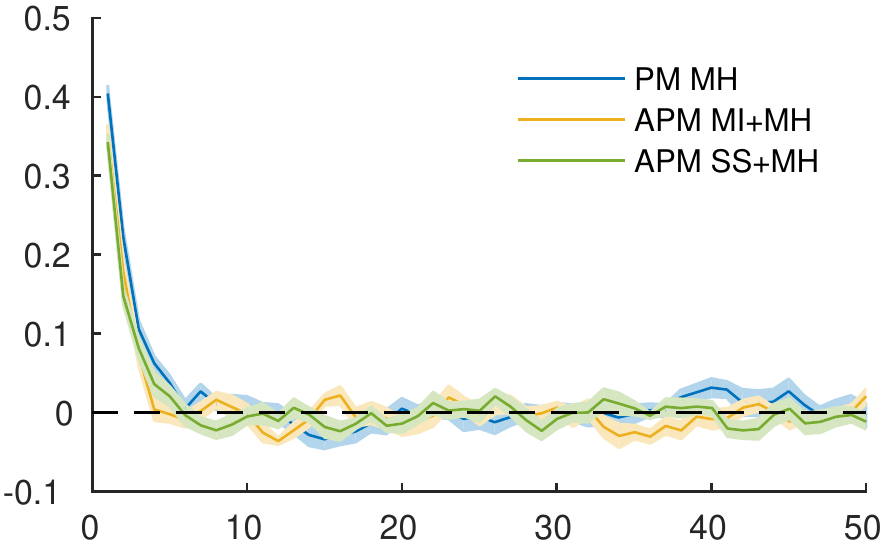}~\label{fig:gpcorr1}}\hspace*\fill\\
\hspace*\fill\subfloat[][$\tau$ given Breast data]{\includegraphics[height=4.0cm]{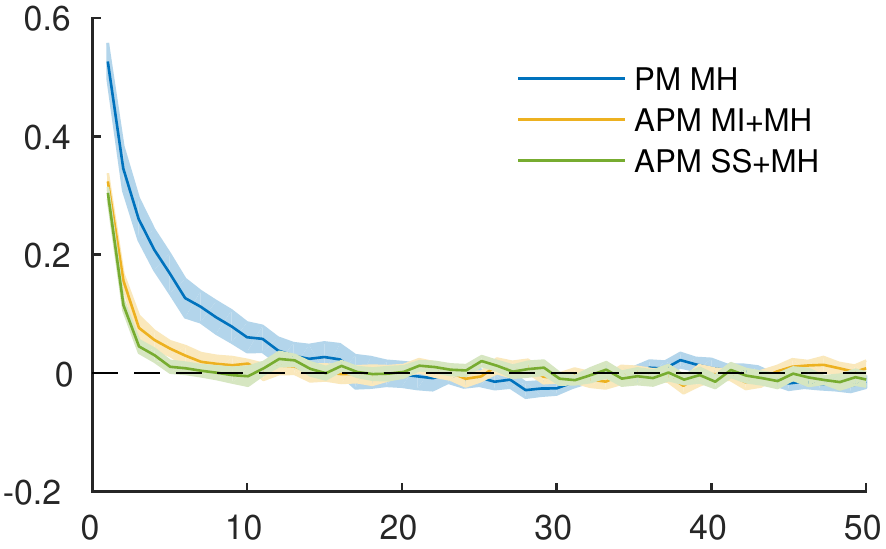}\label{fig:gpcorr2}}\hspace*\fill
\caption{%
Cost-scaled autocorrelations (after thinning to every 10 iterations) for Gaussian Process models.
}
\label{fig:gpcorr}
\end{figure}

We specifically considered Gaussian process parameter inference on two
UCI classification datasets tested by
\citet[\S4.4]{filippone2014}: the Pima data set ($d\te9$, $n\te682$) and the Breast
data set ($d\te8$, $n\te768$). For a full description of the approximation scheme
used in \eqref{eqn:gp_marg_lik_estimator}, parameter prior hyperparameters and
other implementation details please refer to \cite{filippone2014}.

\textbf{Ease of tuning:~} All three methods considered used MH updates for
$\theta$ so we employed the adaptive scheme used in \cite{filippone2014} to
tune the proposals to try to achieve an acceptance
rate between 0.15 and 0.3.
\citet{filippone2014} noted the possibility of chains getting stuck due to large
overestimates of the marginal likelihood, and suggested it could upset the
adaptive process. Therefore a biased but deterministic approximation for the
marginal likelihood was used in the MH accept step during the initial adaptive
phase. Preliminary experiments indicated that their approach worked better, and
so was replicated in our experiments for the PM MH method.

The Auxiliary Pseudo-Marginal (APM) framework is meant to provide a clearer signal for
adapting proposals, as accept/reject decisions are based on a paired comparison
between $\theta$ values with the same random $\bu$ draw.
This intended advantage seems to work in practice:
the final acceptance rate in all 40 of the APM
chains we ran ended up within the desired acceptance rate bounds. For
the standard pseudo-marginal MH case, 2 out of the 20 chains did not
achieve an acceptance rate within the target bounds.

\textbf{Efficiency:~} The sampling efficiencies of the tested auxiliary
pseudo-marginal methods and original pseudo-marginal MH method are summarised
in Table~\ref{tab:gp_param_inf_stats}. Also shown are the Gelman--Rubin
$\hat{R}$ convergence statistics \citep{gelman1992inference}
for each set of 10 chains: values far from
unity would demonstrate failure to converge. Although the diagnostic cannot
prove convergence, the fact that all of the calculated $\hat{R}$ are unity to within
0.01 is at least comforting. The autocorrelations (Figure~\ref{fig:gpcorr}) also
go to zero quickly.

The effective sample sizes for the two auxiliary
pseudo-marginal methods both show significant improvements over
the standard pseudo-marginal MH method, with the gain
being particularly large in the Breast data set where the ESS is more than
doubled and the autocorrelations also show a quicker decay to zero
(Figure~\ref{fig:gpcorr2}).
Traces (not shown) for the PM MH runs on the Breast data set showed
heavy incidence of the chain sticking. These
artefacts are also visible for this data set in \cite{filippone2014}'s plots.
The auxiliary pseudo-marginal methods appear to mix much better with no
extended rejection intervals.

In general the auxiliary pseudo-marginal framework
has some computational overhead over standard pseudo-marginal
due to splitting the update into two steps. For this Gaussian process task however
the dominant computational cost during sampling is in
$\mathcal{O}(n^3)$ matrix decomposition operations, which are only required
when considering a new $\theta$.
Counts for the total number of cubic
operations $N_\text{c.op.}$ across each run were recorded and the mean values
for each method and dataset are shown in Table~\ref{tab:gp_param_inf_stats}.
The computational cost for all three methods was effectively
equivalent (in our implementation the wall-clock time of SS+MH was $\sim6\%$ more per update than MI+MH)\@.

Between the two auxiliary pseudo-marginal methods, applying elliptical slice
sampling to the random draws $\bu$ seems to give a small gain in
sampling efficiency over Metropolis independence sampling or no
significant difference. Falling back to smaller intervals to ensure the
chain always moves appears to help to some extent.

\section{Discussion}

Auxiliary variable interpretations have previously been exploited
in particle MCMC methods, providing new and better update rules
\citeg{chopin2015}.
Since submitting this work, we've been told that the idea of clamping random
number draws in pseudo-marginal methods was first proposed by \citet{lee2010}.
There have also been two independent proposals to update Gaussian random numbers
within arbitrary pseudo-marginal methods \citep{deligiannidis2015,dahlin2015}.
Both of these proposals use the Metropolis proposal mechanism that elliptical
slice sampling generalizes. Without slice-sampling exploration, these methods
will require tuning, and could be less robust. However, Metropolis updates are
simpler to analyse and these related works contain interesting theoretical
analysis of the Markov chains.

The main aim in our presentation of the APM framework is to provide tuning-free
ways to improve pseudo-marginal MCMC methods. Using the framework to clamp the
auxiliary randomness $\bu$ requires no alteration to an estimator's existing
code: one can simply set the random seed it uses. Users can employ their
preferred updates for the target variables $\theta$ (and can now tune their
step-sizes reliably). Using the proposed slice sampling methods to update random
draws $\bu$ requires replacing calls to a random number generator with access to
a Markov chain state. Apart from that, a user can use either our black-box
reflective slice-sampling scheme, or elliptical slice sampling, with no tuning.

In a case where the estimator of the target distribution had low noise
(annealed importance sampling applied to the Ising model), the overhead
of our framework made the wall-clock time for our Markov chain to converge
longer. If prepared to tune free choices, the extra cost of our framework could be reduced by clamping
$\bu$ for several iterations.

However, our biggest concern is making the methods robust. In current
pseudo-marginal methods, it is hard to know what effect noisy estimators
will have on the surrounding Markov chain. \citet{doucet2015} suggest how
much computational effort to spend on reducing noise, but under some
strong assumptions, which don't apply to our examples. In
applications it is difficult to know if an estimator may be heavy tailed,
or behave badly for some parameters~$\theta$. In the Ising example we saw
APM MI+SS suddenly stick after more than a million iterations of apparently
equilibrium behavior. Slice sampling the random choices in an estimator may
cost a little extra, but provides a route for pseudo-marginal chains to
take small steps out of difficulty. This robustness could be the difference
between the Markov chain method working or not.

\subsubsection*{Acknowledgements}
Thanks to
\blind{Luigi Acerbi}
for asking how to apply slice sampling to his pseudo-marginal setup,
\blind{Guido Sanguinetti} for a useful discussion, and
\blind{Chris Sherlock} for pointing out concurrent related work.

This work was supported in part by grants EP/F500385/1 and  BB/F529254/1 for the University of Edinburgh School of Informatics Doctoral Training Centre in Neuroinformatics and Computational Neuroscience (\url{www.anc.ac.uk/dtc}) from the UK Engineering and Physical Sciences Research Council (EPSRC), UK Biotechnology and Biological Sciences Research Council (BBSRC), and the UK Medical Research Council (MRC).

\bibliographystyle{abbrvnat}
\bibliography{bibs}

\end{document}